\documentstyle[aps,manuscript,graphicx]{revtex}
\newcommand{\jpsi}{{J/\psi}}
\newcommand{\psip}{{\psi^\prime}}
\newcommand{\ccbar}{{c\bar c}}
\newcommand{\be}{\begin{eqnarray}}
\newcommand{\ee}{\end{eqnarray}}

\begin{document}

\draft

\title{$J/\psi$ and $\psi^\prime$ total cross sections and formation  times
from data for charmonium suppression
  in $pA$ collisions}

\author{
   Y.B. He,
   J. H\"ufner}
\address{Institut f\"ur Theoretische Physik, Universit\"at Heidelberg,\\
        Philosophenweg 19,
        D-69120 Heidelberg, Germany}
\author{B.Z. Kopeliovich}
\address{Max-Planck Institut f\"ur Kernphysik, Postfach 103980, 69029
   Heidelberg, Germany\\
Joint Institute for Nuclear Research, Dubna, 141980 Moscow Region, Russia}


\maketitle

\begin{abstract}
The recent data for E866
on the $x_F$ dependence for charmonium suppression in $pA$ 
collisions at 800~GeV are analyzed using a time- and 
energy-dependent preformed charmonium 
absorption cross section $\sigma_{\rm abs}^\psi(\tau,\sqrt{s_{\psi N}})$. 
For $\sqrt{s}=10$~GeV the initially ($\tau=0$) produced premeson has an
absorption cross section of $\sigma_{\rm pr}\simeq 3$~mb. 
At the same energy but for $\tau \rightarrow
\infty $ one deduces 
for the total cross sections $\sigma_{tot}^{\jpsi N}=(2.8\pm 0.3)$~mb,
$\sigma_{tot}^{\psip N}=(10.5\pm 3.6) $~mb.
The date are compatible with a formation time $\tau_{1/2}=0.6$~fm/$c$.

\end{abstract}
\pacs{{\bf PACS}: 24.85.=p, 12.38.Mh, 25.75.Dw}
\narrowtext

The formation of a charmonium meson $\jpsi$ or $\psip$ from the originally
produced $\ccbar$ pair via gluon fusion
is a point of discussion and disagreement in the
community which investigates the charmonium suppression in nucleus-nucleus
collisions\cite{review,review2,bla89,kop91,hue96,kha93}. 
What is the nature (color, size) of the premeson when it
travels through nuclear matter? How long does it take to be a fully developed
$\jpsi$ or $\psip$? Are the formation times different for $\jpsi$ and $\psip$,
the $\psip$ being considerably larger than the $\jpsi$? What are the
total 
cross sections $\sigma_{tot}^{\jpsi N}$, $\sigma_{tot}^{\psip N}$,
for the fully developed charmonia on a nucleon? These
questions can be possibly
answered using the recent and still preliminary data of the
E866 collaboration\cite{lei98}. 
The analysis will be given in this paper. 
Since the data are still preliminary, this letter stresses the method of the
analysis, while the numerical results may still undergo certain modifications.

In an
experiment $pA\longrightarrow \psi X$
at 800~GeV,  where $\psi$ stands for $\jpsi$ and $\psip$,
the suppression
\be
S_\psi(x_F)={d\sigma (pA\longrightarrow \jpsi X) \over
A d\sigma (pN\longrightarrow \jpsi X)}
\label{Eq:suppression}
\ee
 has been measured as a function
of $x_F$ for the produced $\psi$ in the range $x_F \gtrsim -0.1$. We will
concentrate in this paper on the interval $-0.1 \lesssim x_F \lesssim 0.25$
where formation time effects are expected to be of particular importance
and where effects of the production time (coherence time) can be neglected.
The effects  of the formation time are determined by the 
Lorentz factor $\gamma(x_F)$ of the charmonium with respect to the
target nucleus. In the experiment pA at 800~GeV, $\gamma$ 
varies between a value of 7 at $x_F=-0.15$ and a value of 47 at
$x_F=+0.15$. For a hypothetical value of the charmonium formation time
$\tau_f=0.5$~fm/$c$ in the c.m.s. 
and for a premeson which is
produced in the middle of a nucleus with radius $R_A=6$~fm, 
a $\psi$ observed with $x_F=-0.15$ is essentially fully developed inside the
nucleus and absorbed with the asymptotic total cross section
$\sigma_{tot}^{\psi N}$, while for 
$x_F=0.15$ the $\psi$ traverses the nucleus in the form of a premeson.

On the other hand, the coherence time $t_c$ 
of charmonium production, which
is the quantum-mechanical uncertainty in the time of production of
the premeson is quite short in the interval of consideration. 
In the light-cone approach 
$t_c$ can be treated as the lifetime
of a fluctuation containing the $\bar cc$. It is given by the energy 
denominator corresponding to such a fluctuation,
\be
t_c=\frac{2\,E_G}{M^2}\ ,
\label{tc.1}
\ee
where $M^2$ is the effective mass of the fluctuation. We assume here
that the $\bar cc$ pair is produced via gluon-gluon fusion accompanied
by gluon radiation which is needed in 
order to end up with a color neutral $\bar cc$ pair. 
It would be incorrect to say that the gluon is radiated {\it after}
its interaction with the target. As usual with bremsstrahlung it
takes place both {\it before} and {\it after} the interaction.
This is a quantum-mechanical uncertainty.
In the light-cone approach $t_c$ corresponds to the lifetime of the whole
fluctuation $G\to\bar c\,c\,G$. This is why Eq.~(\ref{tc.1})
contains the gluon energy $E_G$ in the numerator and in the denominator
the effective mass of the $\bar qqG$  \cite{hkz},
\be
M^2=\frac{k_T^2}{\alpha(1-\alpha)} + 
\frac{M_{\bar cc}^2}{1-\alpha}\ .
\label{tc.2}
\ee
Here $k_T$ and $\alpha$ are the transverse momentum and the fraction of
the initial light-cone momentum carried by the radiated gluon, respectively.
$M_{\bar cc}$ is the effective mass of the $\bar cc$ pair which we assume to 
be of the order of the charmonium mass.
It is important to note that $k_T$ cannot be small since the radiated gluon 
has to resolve the inner structure of the $\bar cc$ pair in order
to make it colorless. Therefore, it should be larger than the inverse
$\bar cc$ separation, $k_T >1/r^{\bar cc}_T \sim m_c$.
At the same time, the fraction $\alpha$ should not be large,
otherwise it will produce a shift in the initial gluon
momentum, which falls off steeply like $(1-x_1)^5$,
resulting in a strong suppression. 
A detailed calculation leads to values $t_c=0.25$~fm/$c$ at $x_F=-0.15$ 
and $t_c=4.1$~fm/$c$ at $x_F=0.15$ in the lab system and $\tau_c=0.04$~fm/$c$
and $\tau_c=0.08$~fm/$c$ in the premeson c.m.s., respectively. Indeed
we will see 
$\tau_c\ll \tau_f$. For values of $0.3<x_F<0.8$ one has $t_c>R$ and the
formation time concept becomes doubtful, another mechanism 
seems to be responsible for the observed suppression. 

In the $x_F$ interval $[-0.1,0.25]$ we model the suppression by assuming that
the premeson is created instantaneously and is absorbed
according to a time and energy dependent
effective absorption cross section 
$\sigma^{\psi N}_{\rm abs} (\tau,\sqrt{s_{\psi N}})$ for 
an experiment, in which a charmonium $\psi$ is observed.
(The effects of the decays $\chi_c\rightarrow J/\psi$ and $\psi^\prime
\rightarrow J/\psi$ will be discussed later.)
The classical expression
\be
\label{Spsi}
S_\psi(x_F)={1 \over A} \int d^2 b \int\limits_{-\infty}^{\infty}
dz\,\rho_A(b,z)\, \exp\left\{
-\int_z^\infty dz^\prime \rho_A(b,z^\prime) \sigma^\psi_{\rm abs} 
\left({z^\prime -z \over \gamma(x_F)}\,,\,\sqrt{s_{\psi N}(x_F)}\right)
\right\}\ ,
\label{class}
\ee
is used 
with the nuclear density distribution $\rho_A(b,z)$ normalized to $A$. 
This expression is an approximation to exact solutions for the
evolution of a wave packet with charmonium quantum numbers
propagating through a medium treated in the
quark \cite{kop91} or in hadronic \cite{hk} representations (see also below).
We believe that Eq.~(\ref{class}) is more intuitive and easier to use
than the exact ones. Our aim is to determine the dependence of 
$\sigma^\jpsi_{\rm abs}(\tau,\sqrt{s_{\psi N}})$
and $\sigma^\psip_{\rm abs}(\tau,\sqrt{s_{\psi N}})$ from a fit of Eq.~(\ref{Spsi}) 
to the E866 data. 

We use the following
expression for $\sigma_{\rm abs}^{\psi N}(\tau,\sqrt{s_{\psi N}})$
with two adjustable parameters $\Sigma_{0}$ and $\Sigma_{\infty}$  
which correspond to the effective absorptive cross section
at short and long times, respectively,
\be 
\label{sigmaabs}
\sigma_{\rm abs}^{\psi N} (\tau,\sqrt{s_{\psi N}})= 
\left[ \Sigma_{\infty} +
(\Sigma_{\infty}-\Sigma_{0})\,\cos (\Delta M\, \tau) \right]
\left({\sqrt{s_{\psi N}} \over 10 {\rm GeV} }\right)^\lambda\ .
\ee
The dependence on the energy $\sqrt{s_{\psi N}}$ is deduced from 
photoproduction experiments ($\gamma p\rightarrow J/\psi p$) 
with $\lambda=0.4$\cite{hue98}
The form of the dependence on the time $\tau$ 
is derived within the following
 quantum mechanical model of two coupled channels for the
evolution of a color neutral $\ccbar $ pair\cite{hue96}\footnote{
We assume that the state 
$\vert \bar cc\rangle$ includes all the Fock components 
with additional gluons and sea quarks, $\vert \bar cc\rangle = 
|\bar cc\rangle_0+|\bar ccG\rangle + ...$}: The
time dependent premeson state $\vert c\bar c (\tau)
\rangle$ with the quantum numbers
of $\jpsi$ and $\psip$ can be expanded in a complete set of hadronic states 
of which we keep only the lowest two ones, $\jpsi$ and $\psip$, and we may
use spinor representation: We denote by
\be
\vert c\bar c(0)\rangle =
{1\over \sqrt{1+R^2}}\,\left(
\begin{array}{c}J/\psi\\R\,\psi'
\end{array}\right)
\label{eq:initial}
\ee
the initially produced superposition of the $J/\psi$
and $\psip$.
The
$|\bar cc\rangle$ wave packet in an interacting
environment is described by the equation,
\be
i\,\frac{|d\,\bar cc(\tau)\rangle}{d\,\tau} =
\left(\widehat Q - {i\over2}\,
\rho_A(b,z=\tau\gamma)\,\gamma\,
\widehat T\right)\,
\vert\bar qq(\tau)\rangle\ ,
\label{eq:evolution}
\ee
with
\be
\hat{Q} = \left(
\begin{array}{cc}
M_{J/\psi} & 0 \\
0          & M_{\psi^\prime}
\end{array} \right),
\qquad \qquad
\widehat T =
\left(\begin{array}{cc}
\sigma_{00} & \sigma_{01}\\
\sigma_{10} & \sigma_{11}
\end{array}\right)
\label{eq:interaction}
\ee
where $\hat{Q}$ is the mass matrix and $\hat{T}$ 
is the interaction amplitude operator containing diagonal
and off-diagonal amplitudes, $\sigma_{00}=\langle \jpsi \vert
\hat{\sigma} \vert \jpsi\rangle $, $\sigma_{01}
=\langle \jpsi \vert
\hat{\sigma} \vert \psip\rangle $, {\it etc.}\cite{hk}.

For a $\vert c\bar c \rangle$ created at
the point $(b,z)$ in the nucleus and observed asymptotically as a $\psi$ one has
the transition probability
\be
\label{Wjpsi}
W_\psi(b,z)={\left|
\langle \psi \vert 
 c\bar c 
({z^\prime -z \over \gamma}) \rangle 
\right|^2_{z'\to\infty}
\over 
\left| \langle \psi \vert c\bar c(0) 
\rangle \right| ^2 }\ ,
\label{eq:absorption}
\ee
where $ \vert c\bar c({z^\prime -z \over \gamma}) \rangle$ 
(it depends also on 
$b$) is the solution
of Eq.~(\ref{eq:evolution}) with the initial state (\ref{eq:initial})
at the point with coordinates $(b,z)$.

Expanding expression (\ref{eq:absorption})
in $\rho_A$ up to the first order we get,
\be
\label{W2}
W_\jpsi=1-\int_z^\infty dz^\prime \rho_A(b,z^\prime) 
\left[\sigma_{00}
+R\sigma_{10}
\cos\left(\Delta M ~{z^\prime -z\over \gamma }\right)\right]
\nonumber
\\
W_\psip=1-\int_z^\infty dz^\prime \rho_A(b,z^\prime) 
\left[\sigma_{11}
+ {1\over R}
\sigma_{01}
\cos\left(\Delta M ~{z^\prime -z\over \gamma }\right)\right]
\ee
with $\Delta M = M_{\psip} - M_{J/\psi}$.

The expressions in square brackets in Eq.~(\ref{W2}) are 
the time dependent effective absorption
cross sections and are of the form assumed in Eq.~(\ref{sigmaabs}). 
These effective cross sections may be positive and negative. In the latter
case one observes an enhanced production if one uses a nuclear target 
\cite{kop91}. 

\begin{figure}[t]
\begin{center}
\includegraphics[height=15cm]{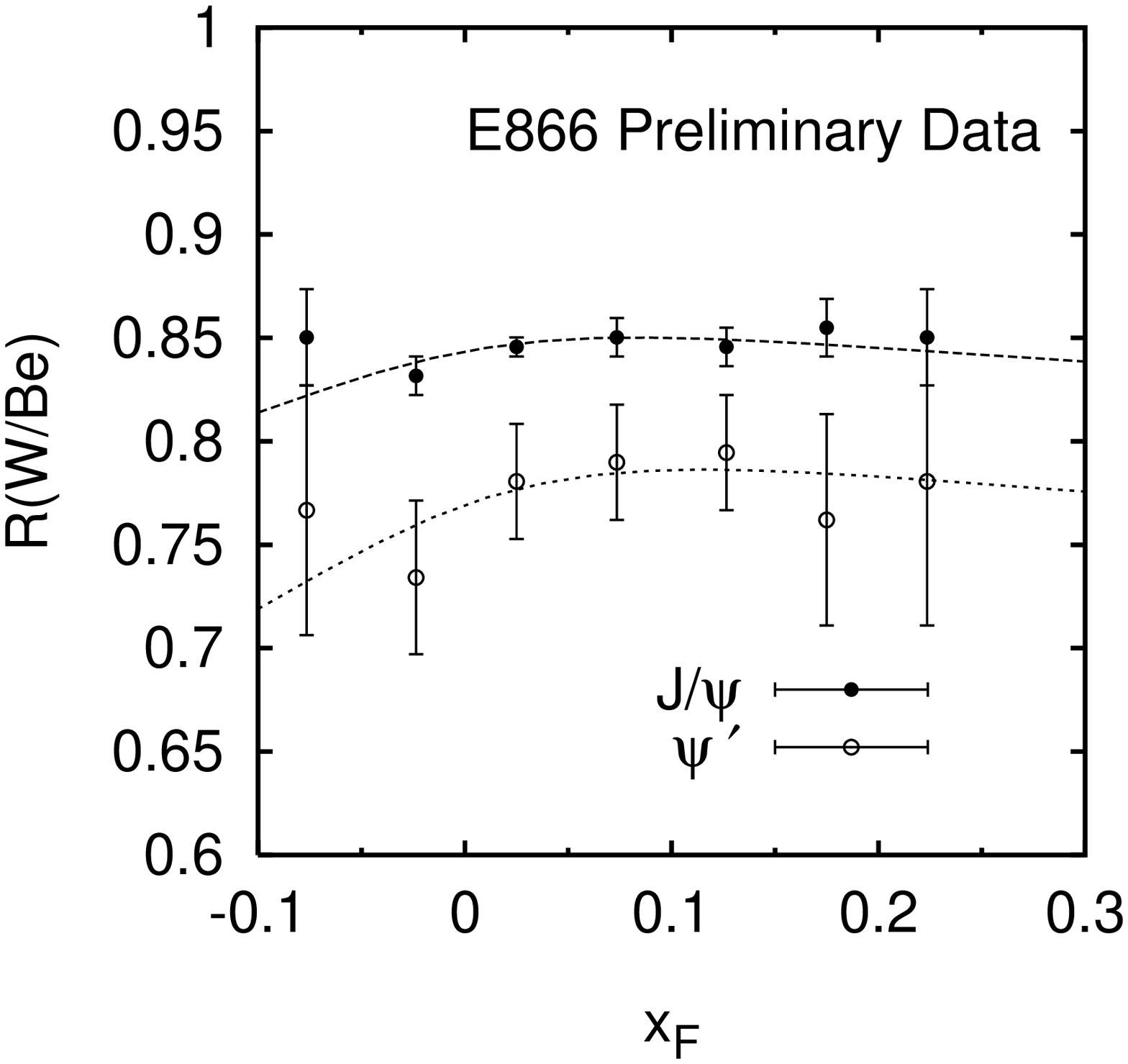}
\caption{
Data from E866 for the ratio of charmonium production in pW and pBe collisions
together with our best fits using 
Eq.~(4).
\label{fig:xFfit}
}
\end{center}
\end{figure}

We have calculated the suppression 
function $S_\psi(x_F)$ in Eq.~(\ref{Spsi})
for Tungsten $A=182$ and Beryllium $A=9$
with a uniform density, and $R=r_0 A^{1/3}$ with $r_0=1.14$~fm. 
The two parameters $\Sigma_0$, and $\Sigma_{\infty}$
for each species of charmonium
have been determined by fitting Eq.~(\ref{Eq:suppression})
to the data of $\jpsi$ and $\psip$ suppression, respectively.
We have used MINUIT-Hesse from CERNLIB.
Fig.~\ref{fig:xFfit} shows the fits. 
The numerical values of the parameters together with their errors and the 
values $\chi^2_{dof}$ as given by
the fit routine are displayed 
in Table 1.

In the following discussion of the results we always take 
$\sqrt{s_{\psi N}}=10$~GeV.

For $\tau \rightarrow \infty$, the oscillating term in the parameterizations
Eq.~(\ref{sigmaabs}) does not contribute in the integral
for the suppression. This is the situation of the fully developed charmonium
and the parameter $\Sigma_{\infty}$ can be identified with the total cross sections,
\be
\sigma_{tot}^{``\jpsi ''N}=(5.0\pm 0.4)~{\rm mb},
\qquad 
\sigma_{tot}^{\psip N} =(10.5\pm 3.6)~{\rm mb}.
\label{sigpsip}
\ee

We have set $J/\psi$ in quotation marks, since the total cross section 
$\sigma_{tot}^{``\jpsi ''N}$ is the one for a situation, where the
observed
$\jpsi$ originates with probability $p_1\simeq 0.6$ 
from directly formed $\jpsi$, 
and
probability $p_2\simeq 0.3$ and $p_3\simeq 0.1$ 
from the decay of $\chi_c$ and $\psip$, respectively.
Thus the total effective absorption cross
section seen in the ``$\jpsi$'' 
channel is a superposition of the contributions of
$\jpsi$, $\chi_c$, and $\psip$. If we correct for this effect, assuming that
total cross sections are proportional to $\langle r^2 \rangle $ one has 
\be
\sigma_{tot}^{\jpsi N} &=&\sigma_{tot}^{``\jpsi ''N} \left[ p_1 + 
{p_2 \langle r^2 \rangle_\chi +
p_3 \langle r^2 \rangle_\psip \over 
 \langle r^2 \rangle_\jpsi } \right]^{-1}
\nonumber
\\
&=& 2.8\pm 0.3~{\rm mb}\ ,
\label{sigpsi}
\ee
where we have used 
$\langle r^2\rangle^{1/2}_{J/\psi}=0.42$~fm,
$\langle r^2\rangle^{1/2}_{\chi}=0.67$~fm,
$\langle r^2\rangle^{1/2}_{\psi^\prime}=0.85$~fm \cite{buc81}. The value in
Eq.~(\ref{sigpsi})
can be compared with $\sigma_{tot}^{\jpsi N}=(3.5 \pm 0.7)$~mb
obtained from an analysis of photoproduction data $\gamma p \rightarrow \jpsi
p$ using the modified vector dominance model\cite{hue98}. 

The ratio of the values from
Eqs.~(\ref{sigpsip}) and (\ref{sigpsi}) gives
\be
{ \sigma_{tot}^{\psip N} \over \sigma_{tot}^{\jpsi N}}
= 3.8\pm 1.3,
\ee
which is close to the value derived from $\psip$ photoproduction\cite{hue98}
 and coincides
with the prediction based on the ratio of the corresponding values of $\langle
r^2 \rangle$ which leads to 4 when the calculations by
Buchm\"uller\cite{buc81} are used. 

Next we discuss the size of the absorption cross sections for a 
premeson directly after creation, {\it i.e.} at $\tau=0$. Then according to
Eq.~(\ref{sigmaabs}), 
the cross sections are related to $\Sigma_0$. Using
the data from Table~1 we have 
\be
\sigma_{\rm pr} =
\left\{ 
\begin{array}{l}
2.7\pm 0.1 ~{\rm mb}\qquad  \jpsi~ {\rm observed}
\nonumber
\\
3.8\pm 0.6 ~{\rm mb}\qquad \psip~ {\rm observed}\ .
\end{array}
\right.
\label{sigcc}
\ee
The two values are rather small compared to the asymptotic cross sections,
supporting the idea that the initial $c\bar c$ system is rather small. The
values in Eq.~(\ref{sigcc}) are equal within the large error bars, although
there is no
compelling reason why they should be the same. 
For instance, one of the origins for a
difference is the admixture of the $\chi_c$-states in the $\jpsi$ channel, 
which at $\tau\rightarrow 0$ cannot be corrected
for easily since the $\langle r^2 \rangle$ law is not applicable.

Within the two channel model 
the time dependence of the effective absorption cross section,
Eq.~(\ref{sigmaabs}) is given by the characteristic time $1/\Delta
M=0.3$~fm/$c$, where $\Delta M$ is the mass difference between the states
$J/\psi$ and $\psi^\prime$. We have not varied this characteristic time since
only few data points are available with rather larger error bars. 
However, since
the parameters $\Sigma_0$ and $\Sigma_\infty$ deduced from the experiment
yield a consistent picture and values of $\chi^2_{dof}< 1$, 
the time dependence derived from
the two channel model may not be unreasonable.
The oscillating dependence of the cross section in Eq.~(\ref{sigmaabs}) makes
the physical interpretation of the characteristic time somewhat
difficult. Since the cross sections always appear under the integral we may
define a formation time $\tau_{1/2}$ by the requirement that the contribution
of the oscillating term is reduced by 50\%, {\it i.e.} we define $\tau_{1/2}$
by 
\be
{1\over \tau_{1/2} } \int_0^{\tau_{1/2}} d\tau \cos \Delta M \tau = {1\over 2}
\ ,
\ee
which happens at $\tau_{1/2}=1.9/\Delta M=0.6$~fm/$c$.

The previous analysis is based exclusively on the data of E866 and 
has yielded the
time and energy dependent absorption cross sections $\sigma_{\rm abs}^{\psi N}
(\tau,\sqrt{s_{\psi N}})$ 
parametrized as in Eq.~(\ref{sigmaabs}). 
We have checked, whether our results are consistent with other data,
for instance, measurements of $\jpsi$ suppression at 200~GeV (NA3) and ratios
of $\psip$ to $\jpsi$ suppression at 450 and 200~GeV (NA38/50).
In the experiment of NA3\cite{NA3}, 
where Pt and $^2$H have been the targets, only four
data points are available for $x_F\lesssim 0.4$, where
formation time effects are 
supposed to be the dominant mechanism. 
We have varied only $\Sigma_\infty$, while setting $\Sigma_0$ at the value
derived from the 800~GeV data.
The results from the
fit are also given in the Table~1 and agree with the results from E866.

Furthermore, we have used the absorption cross sections as deduced from E866
to calculate the ratio of $\psi^\prime / J/\psi$ production on nuclear targets
which has been measured by NA38 and NA50 for several nuclei and energies of
200 and 450~GeV. Results are shown in Fig.~2.
Within the uncertainties induced by 
error bars of the fitted parameters, the data on $\psi^\prime / J/\psi$
suppression can be understood with the time dependent 
absorption cross section.

We summarize: The data of the E866 experiment for the $x_F$ dependence of the
$\jpsi$ and $\psip$ suppressions ($-0.1 \leq x_F \leq 0.25 $)
have been analyzed with the hypothesis that
the variation is entirely due to a time- and energy-dependent 
effective absorption cross section for
a premeson which evolves in time during its passage through the
nucleus. 
The analysis is based on a set of hypotheses and approximations which we list,
but whose accuracies cannot be estimated:
\begin{itemize}
\item We assume that formation time effects dominate the $x_F$-dependence of
  nuclear suppression for $x_F\lesssim 0.25$ where the coherence time
is rather short.
\item We assume that the formation times for $J/\psi$ and $\psi^\prime$
are equal. It is frequently assumed based on the picture of classical
expansion that the $\psip$ needs longer time to  form than the  
$J/\psi$ because of
its larger radius. This might be not true in quantum mechanics.
For instance, for the oscillator potential the formation time
is size independent (the period of oscillation of a pendulum is 
independent of the amplitude).
\item The energy and time dependence of the absorption cross section
  factorizes $\sigma_{\rm
  abs}(\tau,\sqrt{s})=\sigma(\tau)(s/s_0)^\lambda$. The dependence 
of $\lambda$ on the transverse $\bar cc$ separation is rather weak \cite{kp}
and we fix it at $\lambda=0.2$ which follows from $J/\Psi$ photoproduction data.
\item We have assumed one exponential in each channel when calculating the
  suppression function $S_\psi(x_F)$, although the situation is more
  complicated (coupled system of $\jpsi$ and $\psip$, contribution of
  $\chi_c$, {\it etc.}) as is suggested by the
evolution equation (\ref{eq:evolution}) for two coupled channels 
and the exact solution
(\ref{eq:absorption}) which can be found in \cite{hk}.
\end{itemize}
Despite these uncertainties, the analysis has provided a coherent description
of the data from E866, NA3 and NA38/50. We have extracted values for
$\sigma_{\rm tot}^{J/\psi N}$ and $\sigma_{\rm tot}^{\psi^\prime N}$, which
agree with values extracted from photoproduction experiments and follow the
systematics of values of hN total cross sections 
 $\sigma_{\rm tot}^{hN}$ as a function of $\langle r^2 \rangle_h$. For small
 times, $\tau\rightarrow 0$, one finds smaller absorption cross sections, as
 expected, if the premeson is small in size. No statement can be made  about
 the color structure. Although we have not varied the formation time, the fit
 is very good for the formation time $\tau_{1/2}\simeq 0.6$~fm/$c$ derived
 from the two channel model.


After our analysis 
was complete Ref.~\cite{nantes} has appeared in which the same data are
analysed as in this work but with a quite different approach. 
It covers the whole range of $x_F$, and treats the evolution of
the $\bar cc$ pair purely classically.

\acknowledgments
The authors thank G. Garvey for introducing them to these data at an early
state, and to M. Leitch, who communicated the data at various stages of the
analysis.
One of
the authors (J.H.) thanks C. Gerschel for some important suggestions. 
The work has been 
supported in part by a grant from the BMBF
under contract number 06 HD 742.

\begin{figure}[t]
\begin{center}
\includegraphics[height=15cm]{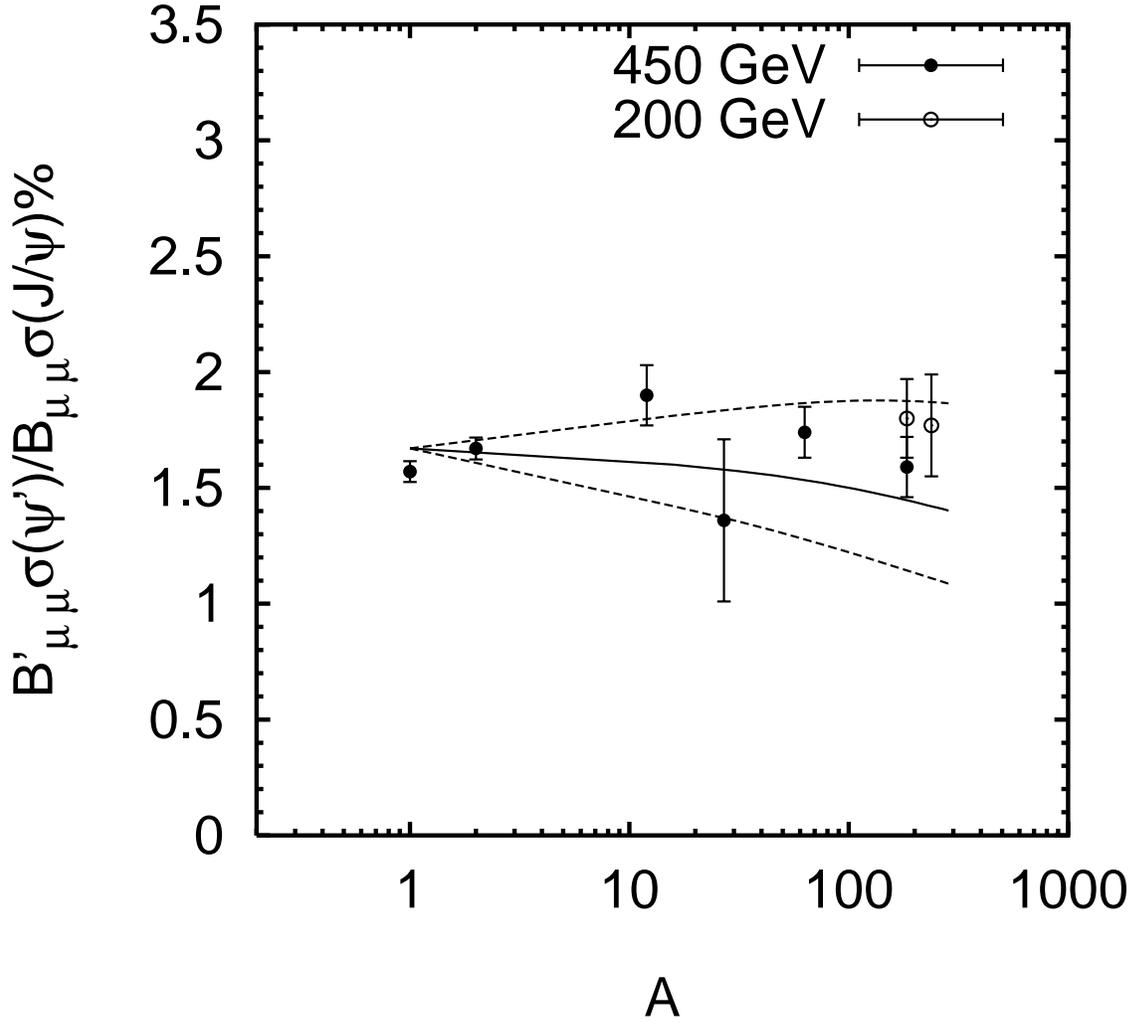}
\caption{Results for the ratio
$B_{\mu \mu}^\prime \sigma(\psip) /B_{\mu \mu} 
\sigma(J/\psi)$
calculated from fitted parameters in Table~1 are compared 
with $pA$ data at and 200 and 450~GeV (data from NA51 and NA38).
The solid curve corresponds to the calculation with parameters at their central
values, while the upper and lower curves characterize the uncertainties
induced by the error bars of the fitted parameters.}
\end{center}
\end{figure}

\begin{table}
\begin{tabular}{c|cccc}
($\psi$) &(exp.) & $\Sigma_\infty$[mb] & $\Sigma_0$[mb] & 
$\chi^2_{dof}$ \\
\hline
$\jpsi$ & (E866)
& $ 5.0 \pm 0.4$ & $2.7\pm 0.1$ & 0.84 \\
$\psip$ & (E866)
& $10.5\pm 3.6$ & $3.8 \pm 0.6$ & 0.28 \\
$\jpsi$ & (NA3)
& $6.8 \pm 1.7$ & $[2.7]$ & 1.6 
\end{tabular}
\vskip 1cm
\caption{Values for the parameters 
$\Sigma_\infty$ and $\Sigma_0$ in the 
parametrization of the absorption cross section 
Eq.~(\ref{sigmaabs}) as obtained from the least square fit to the data.
For the NA3 experiment only the parameter $\Sigma_\infty$ has been fitted,
while the values of $\Sigma_0$ has been set at 2.7.
}
\end{table}

\end{document}